%Final revised version%%
\input harvmac.tex
\parindent=0pt
\parskip=5pt

\def\IR{{\hbox{{\rm I}\kern-.2em\hbox{\rm R}}}}
\def\IB{{\hbox{{\rm I}\kern-.2em\hbox{\rm B}}}}
\def\IN{{\hbox{{\rm I}\kern-.2em\hbox{\rm N}}}}
\def\IC{{\hbox{{\rm I}\kern-.6em\hbox{\bf C}}}}
\def\IZ{{\hbox{{\rm Z}\kern-.4em\hbox{\rm Z}}}}
\noblackbox

\Title{\vbox{\baselineskip12pt
\hbox{CERN-TH/96-62}
\hbox{NSF-ITP-96-14}
\hbox{McGill/96-09}
\hbox{hep-th/9603061}}}
{Entropy of  4D Extremal  Black Holes}

\centerline{ \bf Clifford V. Johnson$^a$, 
Ramzi R.~Khuri$^{b,c}$ and Robert C.~Myers$^c$ }
\bigskip\centerline{$^a${\it Institute for Theoretical Physics, UCSB,
CA~93106, USA }}
\bigskip\centerline{$^b${\it Theory Division, CERN,
 CH-1211, Geneva 23, Switzerland}}
\bigskip\centerline{$^c${\it Physics Department, McGill
University, Montr\'eal, PQ, H3A 2T8 Canada}}
\footnote{}{\tt cvj@itp.ucsb.edu, khuri@nxth04.cern.ch, 
rcm@hep.physics.mcgill.ca}
\vskip1.5cm
\centerline{\bf Abstract}
\vskip0.3cm
\vbox{\narrower\baselineskip=12pt\noindent
We derive the Bekenstein--Hawking entropy formula for four--dimensional
Reissner--Nordstr\"om extremal black holes in type II string
theory. The derivation is performed in two separate (T--dual)
weak coupling pictures. One uses a type IIB bound state problem of D5--
and D1--branes, while the other uses a bound state problem of D0-- and
D4--branes with macroscopic fundamental type IIA strings.  In both
cases, the D--brane systems are also bound to a Kaluza--Klein
monopole, which then yields the four--dimensional black hole at strong
coupling.}

\vskip0.5cm

%%\draft
\Date{\vbox{\baselineskip12pt
\hbox{CERN-TH/96-62}
\hbox{NSF-ITP-96-14}
\hbox{McGill/96-09}
\hbox{March, 1996}}}

\baselineskip13pt

\lref\Bekhawk{J. Bekenstein, Phys. Rev. {\bf D7} (1973) 2333;
Phys. Rev. {\bf D9} (1974) 3292\semi S. W. Hawking,
Comm. Math. Phys. {\bf 43} (1975) 199; Phys. Rev. {\bf D13} (1976)
191.}

\lref\dbranes{J.~Dai, R.~G.~Leigh and J.~Polchinski, Mod.~Phys.~Lett.
{\bf A4} (1989) 2073\semi
P.~Ho\u{r}ava, Phys. Lett. {\bf B231} (1989) 251\semi
R.~G.~Leigh, Mod.~Phys.~Lett. {\bf A4} (1989) 2767\semi
J.~Polchinski, Phys.~Rev.~D50 (1994) 6041, hep-th/9407031.}

\lref\andycumrun{A. Strominger and C. Vafa, {\sl `Microscopic 
Origin of the Bekenstein--Hawking Entropy'}, hep-th/9601029.}

\lref\frank{F. Larsen and F. Wilczek, `Internal Structure of Black Holes',
hep-th/9511064}
\lref\curtjuan{C. G. Callan Jr.  and J. M. Maldacena, {\sl `D--Brane 
Approach to Black Hole Quantum Mechanics'}, hep-th/9602043.}

\lref\garyandy{G. Horowitz and A. Strominger, {\sl `Counting States of
Near--Extremal Black Holes'}, hep-th/9602051.}

\lref\thegang{J. C. Breckenridge, R. C. Myers, A. W. Peet and C. Vafa,
{\sl `D--Branes and Spinning Black Holes'}, hep-th/9602065\semi
J. C. Breckenridge, R. C. Myers, A. W. Peet, A. Strominger and C. Vafa,
in preparation.}
\lref\democrat{P. K.~Townsend, {\sl `P-brane Democracy'},  
hep-th/9507048.}

\lref\hmono{R.~R.~Khuri, Phys. Lett. {\bf B259} (1991) 261;
Nucl. Phys. {\bf B387} (1992) 315, hep-th/9205081.}

\lref\ghs{D.~Garfinkle, G.~T.~Horowitz and A.~Strominger,
Phys. Rev. {\bf D43} (1991) 3140; erratum {\bf D45} (1992) 3888.}

\lref\JKKM{C. V. Johnson, N. Kaloper, R. R. Khuri and R. C. Myers,
{\sl `Is String Theory a Theory of Strings?'}, Phys. Lett. {\bf B368}  
(1996) 71, hep-th/9509070}

\lref\gojoe{J. Polchinski, {\sl `Dirichlet Branes and Ramond--Ramond
Charges in String Theory.'}, Phys. Rev. Lett. {\bf 75} (1995) 4724,
hep-th/9510017.}

\lref\dnotes{J. Polchinski, S. Chaudhuri and C. V. Johnson, {\sl 
`Notes on D--Branes'}, hep-th/9602052.}

\lref\kalloshone{R. Kallosh and B. Kol, {\sl `E(7) Symmetric Area
 of the Black Hole Horizon'}, hep-th/9602014.}

\lref\hetii{C.~M.~Hull and P.~K.~Townsend, {\sl `Unity of Superstring 
Dualities'}, Nucl. Phys. {\bf B438} (1995) 109, hep-th/9410167.}

\lref\monopole{D. J.~Gross and M. J.~Perry, Nucl. Phys. {\bf B226}
(1983) 29\semi
R. D.~Sorkin, Phys. Rev. Lett. {\bf 51} (1983) 87.}

\lref\ko{R. R.~Khuri and T.~Ort\'{\i}n, {\sl `Supersymmetric Black
Holes in $N=8$ Supergravity'}, hep-th/9512177;
{\sl `A Non--Supersymmetric Dyonic
Extreme Reissner--Nordstr\"om Black Hole'}, hep-th/9512178.}

\lref\gibbons{G.W.~Gibbons, Nucl. Phys. {\bf B207}, (1982) 337;
Nucl. Phys. {\bf B298}, (1988) 741.}

\lref\pope{H.~Lu and C.N.~Pope, {\sl `P--brane Solitons in Maximal
Supergravities'}, hep-th/9512012; {\sl `Multi Scalar p--brane
Solitons'}, hep-th/9512153; {\sl `An Approach to the Classification of
p--brane Solitons'}, hep-th/9601089.}

\lref\cvetone{M.~Cveti\u{c} and D.~Youm, {\sl `Four--Dimensional
Supersymmetric Dyonic Black Holes in Eleven--Dimensional Supergravity'}, 
Nucl. Phys. {\bf B453} (1995) 259; {\sl `Dyonic BPS Saturated
Black Holes of Heterotic String on a Six Torus'}, hep-th/9509070\semi
M.~Cveti\u{c} and A.~A.~Tseytlin, {\sl `General Class of
BPS Saturated Dyonic Black Holes as Exact Superstring 
Solutions'}, hep-th/9510097.}

\lref\cvettwo{M.~Cveti\u{c} and A.~A.~Tseytlin, {\sl `Solitonic Strings
and BPS Saturated Dyonic Black Holes'}, hep-th/9512031.}
\lref\cvetthree{M.~Cveti\u{c} and D.~Youm, {\sl `All the Static Spherically
Symmetric Black Holes of Heterotic String on a Six Torus'},
hep-th/9512127; {\sl `Singular BPS Saturated States and Enhanced
Symmetries of Four--Dimensional N=4 Supersymmetric String Vacua'},
Phys. Lett. {\bf 359B} (1995) 87-92.}

\lref\duff{M.J.~Duff, R.R.~Khuri, R.~Minasian and J.~Rahmfeld,
{\sl `New Black Hole, String and Membrane Solutions
 of the Four--Dimensional Heterotic String'}, 
Nucl. Phys. {\bf B418} (1994) 195.}

\lref\dufr{M.J.~Duff and J.~Rahmfeld, {\sl `Massive String States as
Extreme Black Holes'}, Phys. Lett. {\bf B345} (1995) 441.}

\lref\dufp{M.J.~Duff and C.N.~Pope, Nucl. Phys. {\bf B255} (1985) 355.}

\lref\bbj{E.~Bergshoeff, C.M.~Hull and T.~Ort\'{\i}n,
Nucl. Phys. {\bf B451} (1995) 547, hep-th/9504081\semi
K.~Behrndt, E.~Bergshoeff, B.~Janssen, {\sl `Type II Duality
symmetries in Six Dimensions'}, hep-th/9512152.}

\lref\klopv{R.~Kallosh, A.~Linde, T.~Ortin, A.~Peet and
A.~Van Proeyen, Phys. Rev. {\bf D46} (1992) 5278, hep-th/9205027.}

\lref\taubnut{A. H. Taub, Ann. Math. {\bf 53} (1951) 472\semi
E. Newmann, L. Tamburino and T. Unti, J. Math. Phys. {\bf 4} (1963) 915.}
\lref\misner{C. W. Misner, J. Math. Phys. {\bf 4} (1963) 924.}

\lref\rahmfeld{J.~Rahmfeld, {\sl `Extremal Black Holes as Bound
States'}, hep-th/9512089.}

\lref\tsusy{I.~Bakas, {\sl `Space-Time Interpretations of
$S$~Duality and Supersymmetry Violations of $T$~Duality},
Phys. Lett. {\bf 343B} (1995) 103\semi E.~Bergshoeff, R.~Kallosh and
T.~Ort\'{\i}n, {\sl `Duality Versus Supersymmetry and
Compactification'}, Phys. Rev. {\bf D51} (1995) 3009\semi I.~Bakas and
K.~Sfetsos, {\sl `$T$~Duality and World-Sheet Supersymmetry'},
Phys. Lett. {\bf 349B} (1995) 448\semi S.F.~Hassan, {\sl `T-duality
And Non-Local Supersymmetry'}, hep-th/9504148\semi E.~Alvarez,
L.~Alvarez-Gaume and I.~Bakas, {\sl `$T$~Duality and Space-Time
Supersymmetry'}, hep-th/9510028.}

\lref\garyandyjuan{G. Horowitz,
A. Strominger and J. Maldacena, to appear.} 

\lref\vafaone{C. Vafa, {\sl `Instantons on D--Branes'}, hep-th/9512078.}

\lref\bound{E. Witten, {\sl `Bound States of Strings and P--Branes'}, 
hep-th/9510135\semi
M. Li, {\sl `Boundary States of D--Branes and Dy--Strings'}, 
hep-th/9510161\semi
M. Douglas, {\sl `Branes Within Branes'}, hep-th/9512077\semi
C. Vafa, {\sl `Gas of D--Branes and Hagedorn Density of BPS States'}, 
hep-th/9511088\semi
M. Bershadsky, C. Vafa and V. Sadov, {\sl `D--Strings on D--Manifolds'},
hep-th/9510225\semi
A. Sen, {\sl `U--Duality and Intersecting D--Branes'}, hep-th/9511026.}

\lref\ed{E.~Witten, {\sl `String Theory Dynamics in Various Dimensions'}, 
Nucl. Phys. {\bf B443} (1995) 85, hep-th/9503124.}

%%%%%%%%%%%%%%%%%%%%%%%%%%%%%%%%%%%%%%%%%%%%%%%%%%%%%%%%%%%%%%%%%%%%%%%%%%

\newsec{Introduction}
Not long ago, a  long--outstanding goal in theoretical physics
was achieved. The Bekenstein--Hawking formula for black hole entropy
was shown to have a microscopic origin in an explicit string theory
computation. This computation was
performed for extremal five--dimensional
Reissner--Nordstr\"om black holes in ref.\andycumrun\foot{There 
is a  qualitative  proposal for a different
string theory computation of the entropy in the literature. See
refs.\refs{\frank,\cvettwo}.}, and then for the
slightly non--extremal case in refs\refs{\curtjuan,\garyandy}. The
computation was further generalised to five--dimensional rotating
black holes in ref.\thegang.

String theory\foot{We use the term cautiously. It is clear that string
theory is more than a theory of strings, and that the (unknown)
complete theory necessarily admits descriptions in terms of other
extended objects.  Overwhelming evidence for this has been accumulated
over the past year. See, for example, refs.\refs{\democrat,\JKKM}\ for
(pre--D--brane era) discussions of suggestive computations, with
references.  Precise computational power emerged with the
discovery\gojoe\ that the extended objects called D--branes\dbranes\
carry the basic units of Ramond--Ramond charge.  Ref.\dnotes\ contains
a presentation of some of the basic techniques and applications of
D--branes.}\ has long been heralded as a complete theory of quantum
gravity. Over the years, it has taught us that we should think of
black holes as not merely solutions of the background field equations
of string theory, but as being made of strings, in some sense.  The
string folklore is that a black hole solution of the background field
equations is really a condensate of the graviton, described
perturbatively by string theory.  So the problem of computing
properties of the black hole really addresses the problem of
understanding how to describe a black hole in terms of its most basic
constituents, and not simply treating it as a background about which
to perturb. The type of calculation outlined in ref.\andycumrun\ has
begun to make sense of a number of these ideas.

It is a strong coupling problem to compute the entropy of a black
hole. However, the insight gained over the past year into the
structure of strong coupling string theory brings with it the
realisation that we might (in some special cases) be able to compute
the entropy in some weak coupling regime and trust that it may be
successfully extrapolated.  Emboldened by the discovery\gojoe\ of the
relevance of D--brane technology\dbranes\ to the study of
Ramond--Ramond (R-R) charged extended objects in types I and II
superstring theory, the authors of ref.\andycumrun\ demonstrated that
the intuition that D--branes can be thought of as a weak coupling
description of R-R charged black holes can be made precise. 

The moral of the calculation is as follows: Extremal
black holes ({\sl i.e.}, the  strong coupling description)
 with large multiples of the fundamental units of R-R
charge, (and with some large Neveu--Schwarz--Neveu--Schwarz (NS-NS)
charge corresponding to an internal momentum, in the five--dimensional
example of ref.\andycumrun) seem to correspond at weak coupling to
BPS\foot{States satisfying a Bogomol'nyi--Prasad--Sommerfeld 
bound. They are in reduced multiplets of the supersymmetry
algebra. Formulae for their masses and charges are protected by
supersymmetry regardless of the values of the coupling constants in
the theory. So an enumeration of their number is a reliable thing to
extrapolate in coupling space.}\ excitations of bound states of
D--branes carrying these elementary units of R-R charges.  The
degeneracy of these excitations gives a result for the entropy which
precisely matches the entropy which one would deduce from the horizon
area of the extremal black hole using the Bekenstein--Hawking area
law. The fact that the excitations are BPS is the key to complete
confidence in the statement that the extrapolation of this result to
strong coupling makes sense, a procedure which has borne much fruit in
recent times.  Additional exciting results have been presented
recently showing that it is possible (in the right limits) to further
compute the Bekenstein--Hawking entropy from first principles for
non--extremal black holes\refs{\curtjuan,\garyandy}.

All of the examples of this type of computation have been for
five--dimensional black holes. It would be comforting to demonstrate
that computations of the same spirit can be performed in a
four--dimensional setting. Such a demonstration is presented in this
paper.  Our example involves the four--dimensional extremal
Reissner--Nordstr\"om black hole solutions of the Einstein--Maxwell
system. This solution can be embedded into string theory in many
different ways\refs{\klopv\duff\dufr\cvetone\cvettwo\cvetthree\pope{--}\ko}.  
We choose two T--dual embeddings which we study in a six--dimensional
string theory. Our calculation is as follows: Looking at the
five--dimensional subspace $(t,r,\theta,\phi,x^4)$ we see that the
solution looks like a combination of a charged black hole and a
Kaluza--Klein monopole\foot{This is quite distinct from the previous
examples, where at this stage one really has a five--dimensional black
hole.}, where the fifth dimension has been fibred over the angular
coordinates to give a Taub--NUT--like geometry\monopole. In six
dimensions, we see that the solution has a momentum (or winding number
in the T--dual) in the sixth dimension $x^5$. This is very similar to
the situation in the pioneering examples. Once in six dimensions we
see that we have two (T--dual) D--Brane bound state excitation
problems.  One of them is identical to that of ref.\andycumrun.  An
important novelty here is of course that the D--brane composite is
really bound to a Kaluza--Klein monopole, and so we are addressing a
somewhat different bound state problem.  In the strong coupling
solution, the monopole's magnetic charge is essential in maintaining a
constant modulus for the $x^4$ direction, and hence ensure that the
four-dimensional string coupling is constant everywhere.  For the
purposes of the degeneracy counting at weak coupling, the monopole
does not affect the internal structure of the D--brane excitations.

In section~2 we present the two T--dual six--dimensional problems as
uplifted four--dimensional black holes, calling them models `A' and
`B' appropriately after the type II string theories they live in. We
compute and display the charges, horizon area, and then the entropy in
terms of these quantities.  Section~3 briefly describes the weak
coupling bound state problems to which we map the field configurations
of section~2.  We review the degeneracy counting arguments for model B
in section~4 and then show how it translates into the same counting
for the isomorphic bound state problem in model A. The irrelevance of
the monopole to the counting problem is also discussed in section~4.
The entropy computed by explicit enumeration of the degeneracy of BPS
excitations agrees with the Bekenstein--Hawking area law, as
advertised.

\newsec{Four--Dimensional Reissner--Nordstr\"om}
The electrically charged extremal Reissner--Nordstr\"om solution of
the Einstein--Maxwell system:
\eqn\einmax{I\sim \int d^{4}x\ \sqrt{-g}
\left(R 
-{1\over4} F^{2}\right)}
is given by 
\eqn\hetsol{\eqalign{ds^2=&-V^{-2}dt^2+V^2(dr^2+r^2 
d\theta^2+r^2\sin^2\theta d\phi^2)\cr A=&2V^{-1}dt,\qquad V=1+{k\over
r}.}} This solution has a degenerate 
horizon of non--zero area at $r=0$.  It was
first considered in the context of supergravity in
\gibbons\ (see also \dufp). More recently, supersymmetric as well 
as non--supersymmetric
embeddings of the extremal Reissner--Nordstr\"om black hole were found
in \refs{\klopv\duff\dufr\cvetone\cvettwo\cvetthree\pope{--}\ko}.  The
particular embedding we consider preserves $1/4$ of the spacetime
supersymmetries of heterotic string theory compactified on a
six--torus, $T^6$ (see \refs{\cvetone,\ko}\ and references therein).

This solution preserves $1/8$ of the supersymmetries of $N=8$, $D=4$
supergravity, which arises as the low--energy limit of type II string
theory compactified on $T^6$.  In truncating to heterotic string
theory on $T^6$ or type II on $K3\times T^2$, one has to make the
correct sign choice for the fields in accordance with the sign choice
in the truncation from $N=8$ to $N=4$ (or, more precisely, in the
truncation from $N=2$ to $N=1$ in $D=10$). In this manner, one of the
four supersymmetries is preserved in the resultant $N=4$, $D=4$
supergravity limit of these latter compactifications. The opposite
sign choice can still be seen to be supersymmetric, provided the
opposite sign choice is made in truncating from $N=8$ to $N=4$ \ko. In
any case, we choose in this paper the embedding that is supersymmetric
in the appropriate theory, and which under T--duality and other
duality transformations remains supersymmetric with the same amount of
supersymmetry preserved\tsusy.

\subsec{Model B}
The solution may be embedded into six--dimensional type IIB string
theory (compactified on $K3$ for our discussion\foot{We could also
consider a compactification on $T^4$.}) whose action contains the
terms:
\eqn\iibct{\eqalign{
I_{IIB}&={8V_4}\int d^6\!x\, \sqrt{-g}\left[ e^{-2\Phi}
\left(R + 4 (\partial\Phi)^2
-(\partial\sigma)^2-{1\over12}H^2\right)
-{e^{2\sigma}\over12}F_{(3)}^2\right].}}  As is evident from this
action, we are working with the string metric $g$.  The remaining
fields $\sigma$, $\Phi$, $H$ and $F_{(3)}$ are the volume modulus
scalar for $K3$, the six--dimensional dilaton (a linear combination of
the ten--dimensional dilaton and $\sigma$), the 3--form field strength
for the NS-NS 2--form $B$, and the 3--form field strength for the R-R
2--form $A_{(2)}$.  We use units where $\alpha^\prime=1$.  We also
retain $V_4$, the volume of the $K3$ surface.  The six--dimensional
solution is now:
\def\la{{\tilde{\lambda}}}
\def\lt{\lambda}
\eqn\iibsoln{\eqalign{ds^2&=-V^{-2}dt^2+V^2(dr^2+r^2 
d\theta^2+r^2\sin^2\theta d\phi^2)\cr &+k^2(dx^4-\cos\theta\, d\phi)^2
+L^2\left(dx^5+{1\over L}(V^{-1}-1)dt\right)^2;\cr A_{(2)}&={L\over
\lt\la}V^{-1}dt\wedge dx^5-{k^2\over\lt\la}\cos\theta\,d\phi\wedge
dx^4;\cr e^\Phi&=\lt;\qquad e^\sigma=\la; \qquad B=0.}}  where
$V=1+k/r$.

This solution can be seen to be supersymmetric as follows: in \ko\ it
was shown that the truncation of the type IIB theory to an $N=1$,
$D=10$ theory consisting of its NS-NS fields contained the extremal
Reissner-Nordstr\"om black hole as a supersymmetric solution which
preserved a quarter of the remaining supersymmetries. In $D=10$, this
NS-NS solution is related to the ten--dimensional uplift of the
solution \iibsoln\ by interchanging the R-R and NS-NS
two-forms. However, once the 4--form potential generating the
self--dual 5--form field strength of IIB is set to zero, the R-R and
NS-NS 3--form field strengths appear in an identical manner in the
supersymmetry transformations in $D=10$ \bbj.  It then follows that
the ten--dimensional R-R solution is supersymmetric, since the NS-NS
version was already shown to be so \ko. In compactifying back to six
dimensions, the solution \iibsoln\ is then seen to be supersymmetric
and preserves $1/4$ of the spacetime supersymmetries of the $N=2$
theory.

\def\hL{L} 

The metric here is written in the standard form for Kaluza--Klein
reduction.  The first line of the line element $ds^2$ yields the
four--dimensional Reissner--Nordstr\"om in the $(t,r,\theta,\phi)$
subspace. We have chosen the coordinate $x^5$ to have periodicity
$2\pi$, while the $4\pi$ periodicity of $x^4$ is fixed by the fact
that the $(r,\theta,\phi,x^4)$ subspace forms a Euclidean
Taub--NUT--like space\taubnut. The periodicity is chosen to avoid
singularities on $\theta=0$ and $\theta=\pi$. Consequently, the
$(\theta,\phi,x^4)$ subspace has the topology of a 3--sphere\misner.
The five--dimensional subspace $(t,r,\theta,\phi,x^4)$ thus forms
essentially a magnetic Kaluza--Klein monopole \monopole, first found
to be a solution of string theory in \duff\foot{In fact, this
supersymmetric embedding of the extremal Reissner--Nordstr\"om black
hole represents a bound state\refs{\dufr,\cvettwo,\rahmfeld,\ko}\ of two
electric/magnetic dual pairs of dilaton black
holes\refs{\gibbons,\ghs,\klopv}, each in turn representing bound
states of a Kaluza--Klein black hole and an $H$--monopole\hmono. For
our purposes, however, the magnetic Kaluza--Klein monopole plays a
particularly important role.}.  In four dimensions, $g_{44}$ and
$g_{55}$ will play the role of scalar fields, which in this case are
simply the constants $k^2$ and $L^2$. Similarly, $g_{5\mu}/g_{55}$ and
$g_{4\mu}/g_{44}$ play the role of four--dimensional gauge fields.

The solution carries the following six--dimensional conserved
$F_{(3)}$ charges:
\eqn\chargesb{\eqalign{Q_1={V_4\over2\pi^{5/2}}\int_{S^3}e^{2\sigma}\,
{}^*\! F_{(3)}=&{2^3\over\sqrt\pi}V_4{\la\over\lt} k^2;\cr
Q_5={2^3\pi^{3/2}}\int_{S^3}F_{(3)}=&2^7\pi^{7/2}{k^2\over\la\lt}.}}
The off--diagonal metric component $g_{5t}$ indicates that the
solution also carries momentum $P^5$ in the $x^5$
direction. Calculating the total ADM momentum yields
\eqn\momentum{P^5={N\over L}, 
\qquad{\rm where}\,\,\,N=2^8\pi^3 V_4 { k^2L^2\over\lt^2}.}
Calculating the total ADM mass yields:
\eqn\mass{M=2^{10}V_4\pi^3 {k^2L^2\over\lt^2}.}
The extremal black hole has a horizon at $r=0$ which has finite area
given by\foot{We are working with the six--dimensional area here. Of
course, it can be written as the as the product of the
four--dimensional area with the volume of the $x^{4,5}$ space. Note,
however that the correct topology of the horizon is $S^3\times S^1$.
It is more natural to work in terms of six--dimensional quantities,
bearing in mind that the reduction to four--dimensional results is
trivial.}\
\eqn\area{A=2^5\pi^3k^3L.} The Bekenstein--Hawking entropy
formula\Bekhawk\ then states that the entropy of this extremal black
hole is
\eqn\entropy{S={A\over4G_N}=2^{10}
V_4\pi^4{k^3L\over\lt^2},}
where $G_N$ is the effective Newton's constant
$(16\pi G_N)^{-1}=8V_4/\lt^2$.
Note that we can write the entropy in the form $S=2\pi\sqrt{Q_1Q_5N}$.

\subsec{Model A}
\def\hF{{\widehat{F}}}
\def\hA{{\hat{A}}}
We now consider a solution of type IIA string theory compactified on
$K3$ to six dimensions with action:
\eqn\iiact{\eqalign{ I_{IIA}&={8V_4}\int d^6\!x\, \sqrt{-g}\left[ e^{-2\Phi} 
\left(R +4(\partial\Phi)^2
-(\partial\sigma)^2-{1\over12}H^2\right)\right.
\cr
&\qquad\qquad\left.\vphantom{\left({1\over12}H^2\right)}
-{e^{2\sigma}\over4}F_{(2)}^2
-{e^{-2\sigma}\over4}\hF_{(2)}^2
\right],}}
which is:
\def\lp{{\lambda^\prime}}
\def\Lp{{L^\prime}}
\eqn\iiasoln{\eqalign{ds^2&=-V^{-2}dt^2+V^2(dr^2+r^2 
d\theta^2+r^2\sin^2\theta d\phi^2)\cr &+k^2(dx^4-\cos\theta d\phi)^2
+{\Lp^2}(dx^5)^2;\cr e^\Phi&=\lp;\qquad e^\sigma=\la;\cr
B&=\Lp(V^{-1}-1)dt\wedge dx^5;\cr
A_{(1)}&={1\over\lp\la}V^{-1}dt;\qquad
\hA_{(1)}={\la\over\lp}V^{-1}dt,}} where again $V=1+k/r$.  Here
$F_{(2)}$ is the 2--form field strength for the (ten--dimensional) R-R
 vector $A_{(1)}$, while $\hF_{(2)}$ is a 2--form R-R field strength
 for the vector $\hA_{(1)}$. The latter field strength is the
 six--dimensional Hodge dual of the 4--form field strength of the
 (ten--dimensional) type IIA string.  Finally, $B$ is the NS-NS
 2--form potential with field strength $H$.  This solution and the
 previous one
\iibsoln\ are precisely dual under a T--duality
transformation\foot{See for example ref.\bbj\ for explicit details of
how T--duality operates on the background fields in type II theory.}\
in the $x^5$ direction if $\Lp=1/L$ and $\lp=\lt/L$. As noted above
the solutions preserve the same amount of supersymmetry.

The solution has an electric $F_{(2)}$ charge
\eqn\chargesa{Q_0= {V_4\over\pi^{7/2}}\int_{S^3\times
S^1}e^{2\sigma}\,{}^*\!F_{(2)}
={2^3\over\sqrt\pi}V_4{\la\Lp\over\lp}k^2.}  There is also an electric
$\hF_{(2)}$ charge:
\eqn\chargesa{Q_4=4\sqrt\pi\int_{S^3\times
S^1}e^{-2\sigma}\,{}^*\!\hF_{(2)}=2^7\pi^{7/2}{\Lp\over\la\lp}k^2.}
Finally the solution also carries an `electric' charge from the NS-NS
3--form $H$, given by
\eqn\chargesaa{W={16\pi V_4}\int_{S^3}e^{-2\Phi}\,{}^*\!H
=2^8\pi^3 V_4 {k^2\over \lp^2}.} The horizon (still at $r=0$) area is:
\eqn\area{A=2^5\pi^3k^3\Lp.} The Bekenstein--Hawking entropy
formula\Bekhawk\ then states that the entropy of this extremal black
hole is
\eqn\entropy{S={A\over4G_N}=2^{10}\pi^4V_4{k^3\Lp\over\lp^2},}
where $G_N$ is the effective Newton's constant $(16\pi
G_N)^{-1}=8V_4/\lp^2$.  Note that we can write the entropy in the form
$S=2\pi\sqrt{Q_0Q_4W}$.

These results are of course T--dual to the ones derived in the
previous sub--section.

\newsec{Two Bound State Problems}
The next step is to identify the ingredients of the weak coupling
description.  This is done by recalling the result of ref.\gojoe\ that
the basic carriers of R-R charge in type II string theory are
$p$--dimensional extended objects which we shall call here
`D$p$--branes'.  (We reserve the term `D--brane' for the more generic
extended objects, or when we have bound states.)  In ten dimensions,
the world volume of these objects couple to $(p+1)$--form R-R
potentials, and they therefore carry `electric' charges with respect
to a $(p+2)$--form R-R field strength\dnotes.  Therefore, by noting
the amounts of charge that all of the R-R fields in the solutions
\iibsoln\ and
\iiasoln\ have\foot{Note that in the previous sections we have used
normalisations which
ensure that our charges are integer
amounts of the basic units \refs{\gojoe,\dnotes}. 
Thus the R-R charges simply count the number of the corresponding
D--branes in the underlying bound state.
The energetic reader may find it instructive to check that
the analogous formulae yield also the correct integral
quantisation for the
momentum and winding number.}, we can determine the composition of the
D--brane bound states which the solutions become at weak coupling.
The NS-NS charges (associated with $g_{5\mu}$, $g_{4\mu}$
and $B$) then complete our picture of the underlying bound state
configuration. In model A, the electric $B$ charge indicates the
presence of `fundamental' strings winding around $x^5$.

\subsec{Model B}
In the type IIB theory, we have a solution with magnetic charge $Q_5$
and electric charge $Q_0$ with respect to the R-R 3--form field
strength. The fundamental objects which carry these charges in six
dimensions are D1--branes, otherwise known as D--strings.  {}From the
ten--dimensional point of view, there are a number of ways to
construct composite D--strings in compactifying to six dimensions.
Type IIB string theory contains 3--, 5-- and 7--form field strengths
in ten dimensions, for which D1--branes, D3--branes and D5--branes
carry electric charge. In compactifying from ten to six dimensions on
$K3$, the last two extended objects can wrap themselves around the 2--
and 4--cycles of $K3$ and appear as D--strings in six dimensions.
   
The case which we consider here is a compactification in which $Q_5$
D5--branes (which carry magnetic R-R 3--form charge) wrap around the
whole of an internal $K3$, appearing as strings in six dimensions,
forming a composite with $Q_1$ D1--branes (which carry electric R-R
3--form charge).  BPS excitations of such a configuration preserve
$1/4$ of the spacetime supersymmetries in the $N=2$ theory, as the
problem requires. This D--string composite must also have momentum
$P^5=N/\hL$ in the compact $x^5$ direction in order to match the
$g_{5\mu}$ charge of the black hole configuration \iibsoln. This
D--brane composite is the same object as found in ref.\andycumrun.
Finally, we ascertain from the $g_{4\mu}$ charge in the solution
\iibsoln\ the presence of a magnetic Kaluza--Klein monopole. Hence,
the complete configuration is comprised of the D-string composite
bound to a Kaluza--Klein monopole.

\subsec{Model A}
In the (T--dual) type IIA theory, the situation is slightly different.
We have electric NS-NS 3--form charge (winding number) $W$, the
smallest unit of which is carried by fundamental type IIA strings,
together with electric R-R charges $Q_0$ and $Q_4$, which are carried
by D0--branes in six dimensions.  In ten dimensions, our type IIA
compactification has a D4--brane (which carries electric 6--form
charge in ten dimensions) wrapped about the $K3$, appearing as a
D--particle in six dimensions.  The bound state problem is one
involving a D--particle composed of $Q_0$ D0--branes and $Q_4$ wrapped
D4--branes, threaded by\foot{Such configurations of R-R `beads' on a
NS-NS `necklace' have been considered recently in ref.\garyandyjuan\
in a different but not unrelated context.}\ fundamental type IIA
strings winding around the $x^5$ circle with total winding number $W$.
As before, the D--particle is placed in the monopole background, which
provides the extra magnetic charge.  Once again, BPS excitations of
this bound state will preserve 1/4 of the supersymmetries.

\newsec{Microscopic Entropy}
\subsec{Model B}
To evaluate the entropy of the black hole, we need to simply count the
various BPS excitations of the bound states which we discussed in the
previous section. The problem of R-R bound states has received much
attention recently, and much of the counting techniques used in the
black hole computations rely on results derived in
refs\refs{\bound,\vafaone}.

In the problem presented in the type IIB theory the answer is very
familiar, as it has been used several times
now\andycumrun\curtjuan\garyandy.
% Every small child can do it!
Let us set it up as a standard piece of elementary D--brane
calculus\dnotes.  We have $Q_1$ parallel D1--branes and $Q_5$ parallel
D5--branes bound together. This configuration yields the following
decomposition of the spacetime Lorentz group:
\eqn\decompose{SO(1,9) \supset SO(1,1)\otimes SO(4)\otimes SO(4),}
 where the first factor acts along the D--string world sheet
$(t,x^5)$, the third acts in the rest of the D5--brane world--volume
$(x^6,x^7,x^8,x^9)$ and the second in the rest of spacetime
$(x^1,x^2,x^3,x^4)$. In studying the BPS excitations of the bound
states which have the highest degeneracy, we study the (1,5) and (5,1)
open string sector\curtjuan, {\sl i.e.}, oriented strings stretching
between the D1-- and D5--branes\foot{This is true for large $Q_1$ and
$Q_5$}.  There are 2 bosons with NN boundary conditions\foot{Here, `N'
means Neumann, while `D' means Dirichlet.}, 4 with DD and 4 with ND.
Working in the light--cone gauge, these give a zero point energy of
$-$1/12. The NS sector fermions in the four ND directions will be
integer moded. Accordingly, the vacuum energy of the NS fermions
cancels that of the bosons, while the integer moding produces a
degenerate vacuum (like the R sector does in ordinary NN string
theory) forming a spinor under the world--volume $SO(4)$ mentioned above. It
is a boson under the spacetime $SO(1,5)$ Lorentz group.  After the GSO
projection, it is a two--dimensional representation. The R sector
fermions will have half--integer moding in the four ND directions and
thus produce vacuum energy 1/12 again, giving a zero energy degenerate
vacuum, which is a spinor of the $SO(1,5)$, {\sl i.e.}, it is a
spacetime fermion. The GSO projection and a requirement that the state
is left--moving (as we are interested in BPS excitations) reduces the
number of states to two, matching the spacetime bosons from the
previous sector. As (1,5) and (5,1) states are different (we are
considering oriented strings here), we have $4Q_1Q_5$ boson--fermion
ground states.

Our configuration carries momentum $N$ in the $x^5$ direction around
which the D--string is wrapped.  The number of ways, $d(N)$, of
distributing a total momentum $N$ amongst the (1,5) and (5,1) strings
is given by the partition function:
\eqn\partyman{\sum d(N)q^{N}=\left(\prod_{n=1}^\infty 
{1+q^n\over1-q^n}\right)^{4Q_1Q_5}.}  For large $N$, this gives
$d(N)\sim \exp(2\pi\sqrt{Q_1Q_5N})$, and $S=\ln d(N)$ yields precisely
the entropy
\entropy\ we computed for our black hole using the Bekenstein--Hawking
area law, in section~2.

\subsec{Model A}
The counting problem as presented above readily lends itself to
adaptation to other problems. One such is the T--dual type IIA
configuration discussed as the second bound state problem in the
previous section. Here, we have first to consider the D--particle as a
bound state of $Q_0$ D0--branes and $Q_4$ D4--branes, the latter
wrapping around the $K3$ as before. Clearly we do not need to study
this configuration's vacua in any more detail, as we will arrive at
exactly the same conclusions as for the problem above, because they
are T--dual: T--duality exchanges Neumann with Dirichlet boundary
conditions for the $x^5$ direction and so the modings will not change.

We therefore have a bound state D--particle with $4Q_0Q_4$ microstates
in boson--fermion pairs.  However, now we do not have momentum in the
$x^5$ direction. It is the winding number that plays the important
role here. The electric NS-NS 3--form charge of the black hole
solution tells us that there is a large $x^5$ winding number $W$ in
the problem, which is of course carried by the fundamental type IIA
strings. The BPS excitations of the bound state which we want to
consider are therefore those where we have distributed this winding
number $W$ amongst the $4Q_0Q_4$ boson--fermion pairs which we counted
as $(0,4)$ and $(4,0)$ fundamental type IIA strings. We simply ask
that we give these states winding number in as many ways as is
possible to make total winding number $W$. So the configurations of
strings which we considered connecting the constituents of the bound
state are allowed to wind $w$ times around the $x^5$ direction before
connecting between two D--branes. The configurations thus look like
NS-NS `necklaces' made of fundamental winding IIA strings, with a R-R
`bead' (the D--particle composite) somewhere along its length.  The
number of ways, $d(W)$ of distributing the winding $W$ is given by the
partition function:
\eqn\partywoman{\sum d(W)q^{W}=\left(\prod_{w=1}^\infty 
{1+q^w\over1-q^w}\right)^{4Q_0Q_4}.}  This yields the same entropy
previously computed for model B, which is in agreement with the
Bekenstein--Hawking area law.

Once again, we stress that the bound state exists in a Kaluza--Klein
monopole background, whose role at weak coupling was simply to
contribute to the measured charges, but at strong coupling returns us
to the configurations of section~2.

\subsec{The Monopole}
The above discussions (which follow that presented in ref.\curtjuan)
are complementary to the one presented in ref.\andycumrun, where the
same behaviour in the large $N$ (or $W$) limit arises from the
elliptic genus of a sigma model whose target is a symmetric
product\vafaone\ of $K3$.  The $K3$ is assumed small compared to the
circle $x^5$ and the D1--brane components of the D--string bound state
are free to explore the $K3$, about which the (relatively frozen)
D5--brane is wrapped.

It is clear from this point of view that only relative motions of the
D--branes in the $x^5,x^6,x^7,x^8$ and $x^9$ directions are relevant
to the degeneracy counting problem. Excitations in the transverse
directions corresponding to separating the bound state are
irrelevant. The monopole has no structure in the $x^5,x^6,x^7,x^8$ and
$x^9$ directions and hence it only produces a transverse
potential. Thus it will not affect the counting discussions presented
above.

Further note that in isolation the monopole is a soliton and hence has
no intrinsic entropy. Hence the entropy of the total bound state
system arises entirely from the contribution of the D--branes computed
above.

\newsec{Conclusions}
We have demonstrated that the entropy of four--dimensional extremal
Reissner--Nordstr\"om black holes can be computed in essentially the
same way as pioneered in ref.\andycumrun\ for five--dimensional black
holes: Computing in the black hole is a strong coupling string theory
problem, but we can compute the entropy in the weak coupling limit
using D--brane calculus, and extrapolate our results back to the
strong coupling regime, secure in the knowledge that as we have
computed the degeneracy of BPS excitations, our data is protected by
supersymmetry.  A crucial difference between our situation and that of
ref.\andycumrun\ is that our bound states live in a monopole
background. This monopole has no effect on the weak coupling counting
problem, however the extra magnetic charge is essential in producing
the non--singular four--dimensional field configuration.

It is interesting to note that in the five--dimensional cases like
ref.\andycumrun, there are three charges parametrising the black hole,
two R-R and one NS-NS. The precise arrangement of the charges is
essential in producing a nonsingular black hole with a finite area
horizon.  One may observe\ref\joekallosh{S. Ferrara and R. Kallosh, 
{\sl `Supersymmetry and Attractors'}, 
hep-th/9602136\semi J. Polchinski, unpublished.}\foot{We
thank Joe Polchinski for bringing this matter to our attention.} that
all solutions related to the latter under the five-dimensional
U--duality group $E_{6(6)}$ are similarly nonsingular because the
horizon area is related to cubic invariant of $E_{6(6)}$.  In the
present problem, the addition of the monopole charge is essential in
producing a finite area horizon.  The combined arrangement of four
charges must combine in the quartic invariant of the four dimensional
U--duality group, $E_{7(7)}$, (see refs.\refs{\kalloshone,\hetii})
which is conjectured to determine the area of the black hole horizon
\kalloshone.  It would be of some interest to verify explicitly that
the area in the present solutions is given by this $E_{7(7)}$
invariant.

One interesting aspect of the computation was to study how it took on
different forms in a T--dual picture. Of course, the entropy should be
independent of T--duality (and more generally, U--duality\hetii), and
it is pleasing to see how this happens in detail so cleanly.  In
particular with the substitution $\Lp=1/L$ and $\lp=\lt/L$ in the
model A results, we have $Q_0=Q_1$, $Q_4=Q_5$ and $W=N$.  One should
expect that this result will work more generally.  We displayed only a
subset of the class of embeddings of the four--dimensional
Reissner--Nordstr\"om solution into type II string theory. There are
other embeddings which are T--dual to those discussed here, and others
which are not
\refs{\pope,\ko}. One example of the former is to T--dualise model A along 
the $x^4$ direction. The resulting configuration involves D1--branes
wrapping the $x^4$ direction, fundamental strings wrapping the $x^5$
direction and D5--branes wrapping the $K3$ and the $x^4$
direction. There is also a NS-NS solitonic 5--brane wrapping the $K3$
and $x^5$, replacing the role of the monopole.  Some of the latter
embeddings can be obtained, for example, by using string/string
duality\hetii\ed\ to go from type IIA on $K3$ to heterotic string
theory on $T^4$, performing an $O(22,6)$ rotation and then using
string/string duality to return to type IIA.  The class of backgrounds
thus generated contains a rich family of interesting (and presumably
equivalent) bound state problems involving varying amounts of R-R and
NS-NS charged extended objects in diverse backgrounds. It would
certainly be interesting to study more of these, as they would shed
more light on the problem of bound states in string theory, extremal
black holes and perhaps non--extremal black holes.

%%%%%%%%%%%%%%%%%%%%%%%%%%%%%%%%%%%%%%%%%%%%%%%%%%%%%%%%%%%%%%%%%%%%%%%%%%
\bigskip
\medskip
%\vskip0.5truecm

\noindent
{\bf Acknowledgments:}

\noindent
We would like to thank Gary Horowitz, Joe Polchinski and Andy
Strominger for discussions. CVJ was supported in part by the National
Science Foundation under Grant No. PHY94--07194. RCM's research was
supported by NSERC of Canada, Fonds FCAR du Qu\'ebec. RCM would also
like to thank the ITP at Santa Barbara for hospitality while this
research was carried out.  RRK was supported by a World Laboratory
Fellowship.

%\vfill\eject
\listrefs
\bye